\documentclass[
    ,final            
  ]
  {aipproc}

\layoutstyle{6x9}

\usepackage{graphicx} 
\usepackage{amssymb} 
\usepackage{color}

\begin{document}

\title{Status of the LUX Dark Matter Search}

\classification{95.35.+d,}
\keywords      {Dark Matter, LUX}

\author{S. Fiorucci}{
  address={Brown University, Dept. of Physics, 182 Hope St., Providence RI 02912}}

\author{D.~S. Akerib}{
  address={Case Western Reserve University, Dept. of Physics, 10900 Euclid Ave, Cleveland OH 44106}}
\author{S. Bedikian}{
  address={Yale University, Dept. of Physics, 217 Prospect St., New Haven CT 06511}}
\author{A. Bernstein}{
  address={Lawrence Livermore National Laboratory, 7000 East Ave., Livermore CA 94551}}
\author{A. Bolozdynya}{
  address={Moscow Engineering Physics Institute, 31 Kashirskoe shosse, Moscow 115409}}
\author{A. Bradley}{
  address={Case Western Reserve University, Dept. of Physics, 10900 Euclid Ave, Cleveland OH 44106}}
\author{D. Carr}{
  address={Lawrence Livermore National Laboratory, 7000 East Ave., Livermore CA 94551}}
\author{J. Chapman}{
  address={Brown University, Dept. of Physics, 182 Hope St., Providence RI 02912}}
\author{K. Clark}{
  address={Case Western Reserve University, Dept. of Physics, 10900 Euclid Ave, Cleveland OH 44106}}
\author{T. Classen}{
  address={University of California Davis, Dept. of Physics, One Shields Ave., Davis CA 95616}}
\author{A. Curioni}{
  address={Yale University, Dept. of Physics, 217 Prospect St., New Haven CT 06511}}
\author{E. Dahl}{
  address={Case Western Reserve University, Dept. of Physics, 10900 Euclid Ave, Cleveland OH 44106}}
\author{S. Dazeley}{
  address={Lawrence Livermore National Laboratory, 7000 East Ave., Livermore CA 94551}}
\author{L. de~Viveiros}{
  address={Brown University, Dept. of Physics, 182 Hope St., Providence RI 02912}}
\author{E. Druszkiewicz}{
  address={University of Rochester, Dept. of Physics and Astron., Rochester NY 14627}}
\author{R. Gaitskell}{
  address={Brown University, Dept. of Physics, 182 Hope St., Providence RI 02912}}
\author{C. Hall}{
  address={University of Maryland, Dept. of Physics, College Park MD 20742}}
\author{C. Hernandez~Faham}{
  address={Brown University, Dept. of Physics, 182 Hope St., Providence RI 02912}}
\author{B. Holbrook}{
  address={University of California Davis, Dept. of Physics, One Shields Ave., Davis CA 95616}}
\author{L. Kastens}{
  address={Yale University, Dept. of Physics, 217 Prospect St., New Haven CT 06511}}
\author{K. Kazkaz}{
  address={Lawrence Livermore National Laboratory, 7000 East Ave., Livermore CA 94551}}
\author{R. Lander}{
  address={University of California Davis, Dept. of Physics, One Shields Ave., Davis CA 95616}}
\author{K. Lesko}{
address={Lawrence Berkeley National Laboratory, 1 Cyclotron Rd., Berkeley CA 94720}}
\author{D. Malling}{
  address={Brown University, Dept. of Physics, 182 Hope St., Providence RI 02912}}
\author{R. Mannino}{
  address={Texas A \& M University, Dept. of Physics, College Station TX 77843}}
\author{D. McKinsey}{
  address={Yale University, Dept. of Physics, 217 Prospect St., New Haven CT 06511}}
\author{D. Mei}{
  address={University of South Dakota, Dept. of Physics, 414E Clark St., Vermillion SD 57069}}
\author{J. Mock}{
  address={University of California Davis, Dept. of Physics, One Shields Ave., Davis CA 95616}}
\author{J. Nikkel}{
  address={Yale University, Dept. of Physics, 217 Prospect St., New Haven CT 06511}}
\author{P. Phelps}{
  address={Case Western Reserve University, Dept. of Physics, 10900 Euclid Ave, Cleveland OH 44106}}
\author{U. Schroeder}{
  address={University of Rochester, Dept. of Physics and Astron., Rochester NY 14627}}
\author{T. Shutt}{
  address={Case Western Reserve University, Dept. of Physics, 10900 Euclid Ave, Cleveland OH 44106}}
\author{W. Skulski}{
  address={University of Rochester, Dept. of Physics and Astron., Rochester NY 14627}}
\author{P. Sorensen}{
  address={Lawrence Livermore National Laboratory, 7000 East Ave., Livermore CA 94551}}
\author{J. Spaans}{
  address={University of South Dakota, Dept. of Physics, 414E Clark St., Vermillion SD 57069}}
\author{T. Stiegler}{
  address={Texas A \& M University, Dept. of Physics, College Station TX 77843}}
\author{R. Svoboda}{
  address={University of California Davis, Dept. of Physics, One Shields Ave., Davis CA 95616}}
\author{M. Sweany}{
  address={University of California Davis, Dept. of Physics, One Shields Ave., Davis CA 95616}}
\author{J. Thomson}{
  address={University of California Davis, Dept. of Physics, One Shields Ave., Davis CA 95616}}
\author{J. Toke}{
  address={University of Rochester, Dept. of Physics and Astron., Rochester NY 14627}}
\author{M. Tripathi}{
  address={University of California Davis, Dept. of Physics, One Shields Ave., Davis CA 95616}}
\author{N. Walsh}{
  address={University of California Davis, Dept. of Physics, One Shields Ave., Davis CA 95616}}
\author{R. Webb}{
  address={Texas A \& M University, Dept. of Physics, College Station TX 77843}}
\author{J. White}{
  address={Texas A \& M University, Dept. of Physics, College Station TX 77843}}
\author{F. Wolfs}{
  address={University of Rochester, Dept. of Physics and Astron., Rochester NY 14627}}
\author{M. Woods}{
  address={University of California Davis, Dept. of Physics, One Shields Ave., Davis CA 95616}}
\author{C. Zhang}{
  address={University of South Dakota, Dept. of Physics, 414E Clark St., Vermillion SD 57069}}



\begin{abstract}
The Large Underground Xenon (LUX) dark matter search experiment is currently being deployed at
the Homestake Laboratory in South Dakota. We will highlight the main elements of design which
make the experiment a very strong competitor in the field of direct detection, as well as an easily
scalable concept. We will also present its potential reach for supersymmetric dark matter detection,
within various timeframes ranging from 1 year to 5 years or more.
\end{abstract}

\maketitle



\section{The LUX Experiment}
   The LUX Collaboration brings together today nearly 70 scientists from 11 US academic institutions, all with significant experience in the fields of ultra-low background experiments 
\cite{SNO2000,Borexino2008,Eguchi.PhysRevLett.90.021802,Aalseth2005}
 and Dark Matter searches in particular 
\cite{Ahmed:011301,Angle:2007uj,lebedenko-2008}
. The collaboration was formed in 2007 and was fully funded by grants from DOE and NSF in 2008. It has since enjoyed continual growth. The experiment builds upon the now well-established liquid xenon TPC technology \cite{Angle:2007uj,lebedenko-2008}, which provides single electron and photon detection capability, and excellent 3D imaging with centimeter precision. This allows LUX to efficiently reject multiple scatters and to take full advantage of the shelf-shielding of the xenon against external gamma background. Upon interaction of a particle in the detector, a primary scintillation light (S1) is collected on the PMTs. A fraction of the electrons ionized along the recoil track drift toward the top of the detector by application of an electric field, and reach the liquid/gas interface where they are extracted. They are then drifted a short distance in the gas toward a collection anode, which gives rise to a proportional electroluminescence light signal (S2). The combined measurement of S1 and S2 light allows LUX to discriminate between gamma interactions (with the atomic electrons) and neutron, or WIMP, interactions (with the atomic nucleus), the latter giving relatively less ionisation electrons for the same deposited energy. Gamma background rejection with this technique has been shown to be conservatively $\geq 99.5$\% in the energy range of interest for WIMP search \cite{Angle:2007uj}.\\
\\
\textbf{Experimental setup} The LUX detector is a 350 kg Xe target (300 kg fully active in a chamber 59 cm tall $\times$ 49 cm in diameter), viewed by 122 Hamamatsu R8778 2" PMTs.
 The PMTs have been shown to possess a quantum efficiency (QE) consistently better than 30\%, and together with the PTFE reflector panels that delimit the active volume, permit a light collection efficiency in LUX at $\sim$10.0 phe/keVee (Zero Field).

The cryostat holding the xenon target is made of titanium with low radioactivity levels of $<0.4$ mBq/kg in U+Th. These background levels make it a very viable alternative to ultra-pure copper for low-background experiments, while its high strength and low mass make it an excellent choice for vacuum-rated vessels. The inner 100 kg vessel is covered with a thin Cu radiation shield on the outside, and separated from the outter 130 kg vessel by a layer of vacuum. The total detector mass, once assembled and filled with xenon, is $\sim$2.7 tonnes.

The cryogenics involved in the detector design are a significant upgrade from previous, smaller scale liquid xenon detectors. The main source of cooling power is provided by a closed loop of liquid nitrogen condensation and evaporation, through a thermosyphon system. A cold liquid nitrogen bath, sitting outside of the water tank (on top) for safety and practicality reasons, reaches down through a pipe to the detector inside the tank. A primary thermosyphon provides $\sim$1 kW of cooling power, which potentially allows the circulation of xenon in the detector at a rate reaching 1 tonne/day. Two secondary thermosyphons with a smaller power output provide regulation and stability to the system. LUX has also designed a novel heat exchanger, using the cooling power of the pumped, outgoing Xe liquid to cool down the incoming Xe gas in the circulation system. This device has been succesfully deployed in a test cryostat and has displayed $\sim$97\% efficiency at a flow rate of 1/4 tonne Xe/day.

The detector is housed in a 8~m diameter, 6~m tall tank containing 300 tonne of ultra-pure water, which reduces the external gamma background by 9 orders of magnitude (accounting for detector and tank geometry) and makes the neutron background from natural radioactivity essentially negligible.
 The experiment will be hosted at the newly commissioned Sanford Laboratory at Homestake, SD, in the cavern previously used by Nobel laureate Ray Davis for his solar neutrinos experiments
. The 4300 meter-water-equivalent (mwe) rock overburden provides a reduction in the cosmic muon flux by a factor of $4 \times 10^6$ compared to the surface at sea level.\\
\\
\textbf{Expected background and sensitivity} One of the most important advantages of a larger scale TPC is that one gets to take full advantage of the self-shielding properties of liquid xenon, while conserving a significant fiducial mass. The dominant background contribution for LUX, for both gammas and neutrons, comes from the radioactivity of the PMTs (counted levels in U/Th at $9.8 \pm 0.7$ / $2.3 \pm 0.5$ mBq/PMT). Thanks to self-shielding, the Electronic Recoils (ER) background rate is expected to be $260 \times 10^{-6}$ events/keV/kg/day (or 260 $\mu$dru) in a 100 kg fiducial volume, with PMT radioactivity accounting for 90\% of that budget. This corresponds to $\sim$80 ER events expected in 100 kg $\times$ 300 days, in the [5 - 25 keVr] energy range. Another subdominant contribution comes from $^{85}$Kr in the xenon, which has been reduced to <~2~ppt level and amounts for < 10\% of the budget. Other external background sources contribute less than 0.01\% to the gamma background in the fiducial volume. Those rates are before ER/NR discrimination, which can conservatively be set at the 99.5\% level from previous experience with the technology \cite{Angle:2007uj}.

The dominant neutron background contribution comes from ($\alpha$,n) reactions on U and Th in the PMTs. Monte Carlo simulations indicate that the associated rate of single scatters in the 100 kg fiducial volume is expected to be <~0.5~$\mu$dru, \textit{i.e.}, less than 0.15 event in 100 kg $\times$ 300 days. The cosmic neutron background associated with muon-induced spallation in the rock is reduced to negligible levels by the 4300 mwe rock overburden and the instrumentation of the water tank as a Cherenkov veto.

Combined, these extremely low background levels contribute to set the sensitivity reach of the experiment at a WIMP-nucleon cross-section of $7 \times 10^{-46}$ cm$^2$ for a WIMP mass of 100 GeV/c$^2$, and a total fiducial exposure of 100 kg $\times$ 300 days \cite{GaitskellDMTools}. Underground running of the detector is scheduled for summer 2010.

\section{Status of Operations}

\textbf{LUX 0.1 @ CWRU} Since late 2007, development work for the LUX detector has been underway at Case Western Reserve University. The large-scale cryogenic infrastructure and Xe purification systems for LUX have been developed through the ``LUX 0.1'' program, a full-sized prototype cryostat made from stainless steel and containing $\sim$60 kg of Xe, and a small, 4 PMT, 5 cm drift detector housed in a 270 kg Al displacer block. Notable achievements of this program are: successful implementation of a high-flow xenon circulation/purification system, up to 250~kg/d; cooldown and stable operation at liquid xenon temperature with a half of the full LUX cold mass; testing and characterization of all 122 LUX PMTs; acquisition of S1 and S2 light signals for $^{57}$Co 122 keV gammas, greatly advancing the finalization of the DAQ system. The purification has proven to work exceedingly well, reaching a measured >1 m electron drift length after less than 3 days of recirculation.\\
\\
\textbf{LUX 1.0 @ Homestake Surface Facility} At the end of September 2009, the experiment will be moved to a refurbished Surface Facility at Homestake. The detector will use its final design inside the titanium cryostat. The facility has been designed to replicate as closely as possible the layout of the underground laboratory, including a reduced size 3~m diameter water tank to help shield against some of the overwhelming background present at the surface, and a class 1,000 clean room for assembly of the detector. With reduced PMT gains and other measures taken to avoid saturation, the detector is expected to provide limited data taking capability --enough to test xenon purity and S1 trigger efficiency. The setup will be invaluable in debugging the detector internals, cryogenics, electronics and mechanical connections prior to underground deployment.\\
\\
\textbf{LUX 1.0 @ Homestake Davis Laboratory} Access to the 4850-ft level of the Homestake Mine, where the Davis laboratory is located, was made possible again on May 13$^{th}$, 2009 after more than a year of pumping water out of the flooded mine. Dismantling of the remaining infrastructure in the cavern began at the end of July, and LUX will be able to move in by the end of Spring 2010. The new Davis Laboratory will include an 8~m diameter water shield, a class 1,000 clean room and a control room, as well as two germanium counting facility rooms. The whole cavern will be a class 100,000 clean environment.

\section{Outlook and the LUX-ZEPLIN program}
Beyond LUX, a new collaboration called LUX-ZEPLIN or LZ has already been formed by members from LUX and ZEPLIN-III, along with new US collaborators. The ambition of this international team is two-fold. In a first instance, a 1.5 tonne detector will be constructed to replace the LUX detector in the Davis laboratory, with the goal of stable operation in 2012. For a relatively modest investment in R\&D and equipment, this detector has the potential to improve the WIMP sensitivity reach by a factor of 50 compared to LUX, by the end of 2013 (see projections at \cite{GaitskellDMTools}). 

In the longer term, a 15-20 tonne scale detector is the ultimate goal of the LZ collaboration. Such an experiment would be part of the Deep Underground Science and Engineering Laboratory (DUSEL) project at Homestake, and would start construction as soon as 2013 in a new laboratory space at the 7400-ft level. Preliminary studies indicate that 20 tonnes of liquid xenon is a natural limit for this technology, as irreducible cosmic backgrounds from p-p solar neutrinos (for electron recoils), $^8$B solar neutrinos, atmospheric neutrinos and diffuse supernova background (for nuclear recoils via coherent neutrino scattering) all become  significant in order to reach cross-sections of $10^{-48}$ cm$^2$, which is the potential sensitivity of a 20 tonne detector in three years of data taking.

\bibliographystyle{aipprocl} 
\bibliography{LUX_BibTeX_v1}

\end{document}